\documentclass[12pt]{article}
\usepackage[mathscr]{eucal}
\usepackage{amssymb,latexsym}
\usepackage{verbatim}
\usepackage{amsmath}
\usepackage{amsthm}
\usepackage{enumerate}

\newtheorem{thm}{Theorem}[section]

\newtheorem{prop}[thm]{Proposition}

\newcommand\calV{{\mathcal{V}}}
\newcommand\calY{{\mathcal{Y}}}

\newcommand\calT{{\mathcal{T}}}

\renewcommand\l{\lambda}

\newcommand\Om{\Omega}

\renewcommand\S{\Sigma}
\renewcommand\Om{\Omega}
\newcommand\s{\sigma}
\renewcommand\d{\partial}

\newcommand\f{\phi}

\newcommand\D{\nabla}

\renewcommand\div{{\rm div}}

\newcommand\la{\langle}
\newcommand\ra{\rangle}

\newcommand\ric{{\rm Ric}}
\renewcommand\l{\lambda}
\newcommand\g{\gamma}
\newcommand\8{\infty}

\renewcommand\th{\theta}

\newcommand\<{\la}
\renewcommand\>{\ra}

\newcommand\beq{\begin{equation}}
\newcommand\eeq{\end{equation}}
\newcommand\ben{\begin{enumerate}}
\newcommand\een{\end{enumerate}}
\newcommand\bit{\begin{itemize}}
\newcommand\eit{\end{itemize}}

\newcounter{mnotecount}[section]

\title{Some remarks on the size of bodies and black holes}

\author{Gregory J. Galloway 
\thanks{email: galloway@math.miami.edu}
 \\Department of Mathematics \\ University of Miami, U.S.A. \\ \\ Niall \'O Murchadha
  \thanks{email: niall@ucc.ie} \\
 Physics Department \\ University College, Cork, Ireland
}
\begin{document}
\date{}
\maketitle
\vspace{.2in}

\begin{abstract}  We consider the application of stable marginally outer trapped
surfaces to problems concerning the size of material bodies and the area of black holes.
The results presented extend to general initial data sets $(V,g,K)$ previous results
assuming either maximal (${\rm tr}_g K = 0$) or time-symmetric ($K = 0$) initial
data.
\end{abstract}

\section{Introduction}  Let $\S$ be a co-dimension two  spacelike submanifold 
of a spacetime $M$.  Under suitable orientation assumptions, there exists two families
of future directed null geodesics issuing orthogonally from $\S$.  If one of the families has vanishing expansion along $\S$ then
$\S$ is called a marginally outer trapped surface (or an apparent horizon).  The notion of a
marginally outer trapped surface  (MOTS) was introduced early on in the development of the
theory of black holes, and plays a fundamental role in quasi-local descriptions of 
black holes; see e.g.,  \cite{AK}.  MOTSs arose in a more purely mathematical context 
in the work of Schoen and Yau \cite{SY2} concerning the existence
of solutions to the Jang equation, in connection with their proof
of positivity of mass.   

Mathematically, MOTSs may be viewed as spacetime
analogues of minimal surfaces in Riemannian manifolds.  Despite the
absence of a variational characterization for MOTSs
like that for minimal surfaces,  MOTSs have recently
been shown to satisfy a number of analogous properties; see for example,
\cite{AMS0,AMS,AM1,AM2, AG, E, GS}.   Of importance to many of these developments
is the fact, first discussed in \cite{AMS0}, that MOTSs admit a notion of stability analogous, in the analytic sense, to that of minimal 
surfaces (cf., Section 2).  

In this paper we consider  applications of stable MOTSs to two problems
in general relativity.   In Section 3 we address the issue of how the size of a
material body tends to be restricted by the amount of matter contained within it.  
More specifically, we consider an extension of a result of Schoen and Yau \cite{SY}
concerning the size of material bodies to nonmaximal initial data sets. 
In Section 4 we discuss a higher dimensional version of  the lower area (entropy) bounds
obtained by Gibbons \cite{Gi} and Woolgar \cite{Wo}  for  ``topological black holes" which
can arise in spacetimes with negative cosmological constant.   This extends a
result in \cite{CG} to the general nontime-symmetric setting.  We defer 
further discussion of these problems until Sections 3 and 4.   In the next section we 
present some basic background material on MOTSs relevant to our needs.

\section{Marginally outer trapped surfaces}

We recall here some  basic definitions and facts about marginally outer
trapped surfaces.   We refer the reader to \cite{AMS, AM1, GS,G} for further
details.  

Let $V$ be a spacelike hypersurface in an $n+1$ dimensional, $n \ge 3$, spacetime $(M,g_M)$. 
Let $g = \<\,,\,\>$ and $K$ denote the induced metric and second fundamental form 
of $V$, respectively.  To set sign conventions, for vectors $X,Y \in T_pV$, $K$ is defined
as, $K(X,Y) = \<\D_X u,Y\>$, where $\D$ is the Levi-Civita connection of $M$ and $u$ is the future directed timelike unit vector field to $V$.  Note that we are using the `Wald',
rather than the `ADM/MTW', convention for the extrinsic curvature, i.e., positive ${\rm tr}\,K$ implies expansion.

Let $\S$ be a smooth compact hypersurface in $V$, perhaps with boundary $\d\S$, 
and assume $\S$ is two-sided  in $V$.   Then $\S$ admits a smooth unit normal field
$\nu$ in $V$, unique up to sign.  By convention, refer to such a choice as outward pointing. 
Then $l = u+\nu$ is a future directed outward pointing null normal vector field along $\S$, unique
up to positive scaling.   

The null second fundamental form of $\S$ with respect to $l$ is, for each $p \in \S$,
the bilinear form defined by,
\beq
\chi : T_p\S \times T_p\S \to \Bbb R , \qquad \chi(X,Y) = g_M(\D_Xl, Y) \,.
\eeq
The null expansion  $\th$ of $\S$ with respect to $l$  is obtained by tracing the
null second fundamental form, 
$\theta = {\rm tr}_h \chi = h^{AB}\chi_{AB} = {\rm div}\,_{\S} l$, where
$h$ is the induced metric on $\S$.  In terms of the initial data $(V,g,K)$,
$\th = {\rm tr}_h K + H$, where $H$ is the mean curvature of $\S$ within
$V$.  It is well known that the sign of $\th$ is invariant under positive scaling 
of the null vector field $l$.

If $\th$ vanishes then $\S$ is called a marginally outer trapped surface (MOTS).   As mentioned in the introduction, MOTSs may be viewed as  spacetime analogues of minimal
surfaces in Riemannian geometry.  In fact in the time-symmetric case ($K=0$)
a MOTS $\S$ is simply a minimal surface in $V$.  Of particular relevance
for us is the fact that MOTSs admit a notion of stability analogous to that of minimal 
surfaces, as we now discuss.

Let $\S$ be a MOTS in $V$ with outward unit normal $\nu$.   We consider
variations $t \to \S_t$   of  $\S = \S_0$, 
$- \epsilon < t  <  \epsilon,$ with variation vector field 
$\calV = \left . \frac{\d}{\d t}\right |_{t=0} = \phi \nu$,  $\phi \in C_0^{\infty}(\S)$, where
$C_0^{\infty}(\S)$ denotes the space of  smooth functions on $\S$ that vanish on the boundary of $\S$, if there is one.  
 Let $\th(t)$ denote
the null expansion of $\S_t$ with respect to $l_t = u + \nu_t$, where $u$ is the future directed timelike unit normal to $V$ and $\nu_t$ is the
outer unit normal  to $\S_t$ in $V$.   A computation shows,
\beq\label{op}
\left . \frac{\d\th}{\d t} \right |_{t=0}   =   L(\f) \nonumber 
\eeq
where $L : C_0^{\infty}(\S) \to C_0^{\infty}(\S)$ is the operator,
\beq
L(\phi)  = -\triangle \phi + \<X,\D\phi\>  + 
\left( \frac12 S - (\mu + \<J,\nu\>) - \frac12 |\chi|^2+{\rm div}\, X - |X|^2 \right)\phi \,.
\eeq
In the above, $S$ is the scalar curvature of $\S$, $\mu = G(u,u)$, where $G = \ric_M -\frac12R_Mg_M$ is  the Einstein tensor of spacetime, $J$ is the vector field on $V$ dual to the one form $G(u,\cdot)$, and $X$ is the vector field  on $\S$  defined by taking the tangential part of $\D_{\nu}u$ along $\S$.
 In terms of initial data, the Gauss-Codazzi equations
imply, $\mu = \frac12\left(S_V + ({\rm tr}\,K)^2 - |K|^2\right)$  and 
$J = (\div K)^{\sharp} - \D({\rm tr}\, K)$.   

In the time-symmetric case,  $\th$ becomes the mean curvature $H$, the vector field
$X$ vanishes and $L$ reduces to the classical stability operator of minimal surface theory.
In analogy with the minimal surface case, we refer to $L$ in \eqref{op} as the stability
operator associated with variations in the null expansion $\th$.  Although in general $L$ is not self-adjoint,  its principal eigenvalue\footnote{If $\S$ has nonempty boundary, 
we mean the principal Dirichlet eigenvalue.} (eigenvalue with smallest real part) 
$\l_1(L)$ is real.   Moreover there exists
an associated eigenfunction $\phi$ which is positive on $\S \setminus \d\S$. 
Continuing the analogy with the minimal surface case,
we say that a MOTS is stable provided $\l_1(L) \ge 0$.  (In the minimal surface case
this is equivalent to the second variation of area being nonnegative.) Note that if $\phi$ is positive, we are moving `outwards' from the MOTS $\S$, and if  there are no outer trapped surfaces outside of $\S$,  then there shall exist no positive $\phi$ for which $L(\phi) < 0$.  It follows in this case that $\S$ is stable \cite{AMS,AM1,G}.

As it turns out, stable MOTSs share a number of properties in common with  minimal surfaces.  This sometimes  depends on the following fact.   Consider the 
``symmetrized" operator
$L_0: C_0^{\infty}(\S) \to C_0^{\infty}(\S)$,
\beq\label{symop}
L_0(\phi)  = -\triangle \phi  + \left( \frac12 S - (\mu + \<J,\nu\>) - \frac12 |\chi|^2\right)\phi \,.
\eeq
formally obtained  by  setting $X= 0$ in \eqref{op}.   Then arguments in \cite{GS}
show the following (see also \cite{AMS}, \cite{G}).

\begin{prop}\label{eigen}  
$\l_1(L_0) \ge \l_1(L)$.
\end{prop}

We will say that a  MOTS is symmetric-stable if  $\l_1(L_0) \ge 0$; hence ``stable"
implies ``symmetric-stable". 

\section{On the size of material bodies}

In this section we restrict attention to four dimensional spacetimes $M$,
and hence three dimensional initial date sets $(V,g,K)$, $\dim V = 3$. 

It is a long held view in general relativity that  the size of a material body
is limited by the amount of matter contained within it.  There are several
precise results in the literature supporting this point of view.   In \cite{FG},
it was shown, roughly, that the size of a stationary fluid body is bound by the reciprocal
of the difference of the density and rotation of the fluid.  In this case ``size" refers to the
radius of the largest distance ball contained in the body.  

More closely related to the considerations of the present paper is the result of Schoen and Yau \cite{SY} which asserts that
for a maximal (${\rm tr}\, K = 0$) initial data set $(V,g,K)$, the size of a body 
$\Omega \subset V$ is bound  by the reciprocal of the square root of the
minimum  of the energy density $\mu$ on $\Om$.
In this case ``size" refers to the radius of the largest tubular neighborhood in $\Omega$
of a loop contractible in $\Om$ but not contractible in the tubular neighborhood.
As was discussed in \cite{OM1},
this notion of size can be replaced by a notion based on the size of the largest
stable minimal surface contained in $\Omega$.\footnote{This is formulated most simply
when $\Omega$ is bounded and {\it mean convex}, meaning that the boundary
of $\Omega$ has mean curvature $H > 0$.    Then geometric measure theory
guarantees the existence of many smooth least area surfaces contained in $\Omega$.} As argued there, this in general
gives a larger measure of the size of a body, but must still satisfy the same 
Schoen-Yau bound.   The aim of this section is to observe that a similar result holds
without the maximality assumption if one replaces minimal surfaces with MOTS.

Let $V$ be a $3$-dimensional spacelike hypersurface, which gives rise to the
initial data set $(V,g,K)$, as in Section 2.  Consider a {\it body} in $V$ by which we mean a
connected open set $\Om \subset V$ with smooth boundary $\d\Om$. 
 We  describe a precise
measure of the size of $\Omega$ in terms of MOTSs contained within $\Omega$.  
Let  $\S$ be a compact connected surface with boundary $\d\S$ contained in $\Omega$. Let $x$ be a
point in $\S$ furthest from $\d\S$ in $\Omega$, i.e., $x$ satisfies,  $d_{\Omega}(x, \d\S) 
= sup_{y \in \S}\, d_{\Omega}(y, \d\S)$, where $d_{\Omega}$ is distance measured
within $\Omega$.  Then the (ambient) radius of $\S$, $R(\S)$, is defined as $R(\S) =  d_{\Omega}(x, \d\S)$.  

We then define the radius of $\Om$, $R(\Om)$ as follows,
\beq
R(\Om) = \sup_{\S}  R(\S)  \,,
\eeq
where the sup is taken over all compact connected symmetric-stable MOTSs with boundary contained
in $\Om$.    Now this can only be a reasonable measure of the size of
$\Om$ if there are a plentiful number of large symmetric-stable MOTSs contained
in $\Omega$.   But in fact a recent result of Eichmair \cite{E} guarantees the existence of such MOTS, subject to  a natural convexity condition on the body $\Om$.  
We say that $\Om$
is a {\it null mean convex body} provided its boundary $\d\Om$ has positive 
outward null expansion, $\th_+ > 0$, and negative inward null expansion, $\th_-< 0$.
The following is an immediate consequence of Theorem 5.1 in \cite{E}.

\begin{thm} Let $\Omega$ be a relatively compact null mean convex body, with connected boundary, in the
$3$-dimensional initial data set $(V,g,K)$.   Let $\s$ be a closed curve on $\d\Om$
that separates $\d\Om$ into two connected components.   Then there exists a smooth 
symmetric-stable MOTS $\S$ contained in $\Om$ with boundary~$\s$.
\end{thm}

The fact that $\S$ is symmetric-stable follows from a straight forward modification of arguments in \cite[p. 254]{SY2};  see also the discussion   at the end of Section 4 in \cite{E}. 
In fact, a variation of the arguments in \cite[Section 4]{AM2},  may well imply that the MOTS 
$\S$ constructed in Eichmair's theorem  is actually stable.  If that were the case, then
$R(\Omega)$ could be defined in terms of stable, rather than symmetric-stable, MOTS, which we believe would be conceptually preferable.  

We now state our basic result about the size of bodies.

\begin{thm}\label{bound}  Let $\Om$ be a body in the initial data set $(V,g,K)$, and suppose 
there exists $c> 0$ such that $\mu - |J| \ge c$ on $\Om$.  Then, 
\beq
R(\Om) \le  \frac{2\pi}{\sqrt {3}}\cdot \frac1{\sqrt{c}} \,.
\eeq
\end{thm}  

\proof  The proof is similar to the proof  of Theorem 1 in \cite{SY}.   The latter
follows essentially as a special case of the more general Proposition 1 in \cite{SY}.
For the convenience of the reader we present here a simple direct proof of Theorem
\ref{bound}, which involves  a variation of the arguments in \cite{SY}.

Let $\S$ be a symmetric-stable MOTS with boundary $\d\S$ in $\Om$; hence
$\l_1 = \l_1(L_0) \ge~0$.    Choose an associated eigenfunction $\psi$ such that
$\psi > 0$ on $\S\setminus \d\S$.  In fact, by perturbing the boundary $\d\S$ ever so slightly
into $\S$, we may assume without loss of generality that $\psi > 0$ on $\S$.   Substituting
$\phi = \psi$ into Equation \eqref{symop}, we obtain,
\beq\label{laplace}
\triangle \psi = - (\mu + \<J,\nu\>  +  \frac12 |\chi|^2 + \l_1 - \kappa) \psi
\eeq
where $\kappa= \frac12 S$ is the Gaussian curvature of $\S$ in the induced metric $h$.

Now consider $\S$ in the conformally related metric $\tilde h = \psi h$.   The Gaussian
curvature of $(\S, \tilde h)$ is related to the Gaussian curvature of $(\S, h)$ by,
\beq\label{relate}
\tilde \kappa = \psi^{-2} \kappa - \psi^{-3} \triangle \psi + \psi^{-4} |\psi|^2 \,.
\eeq
Combining \eqref{laplace} and \eqref{relate} we obtain,
\beq\label{gauss}
\tilde \kappa = \psi^{-2}(Q + \psi^{-2} |\D\psi|^2)   \,,
\eeq
where, 
\beq\label{Q}
Q = \mu + \<J,\nu\> + \frac12 |\chi|^2 + \l_1  \,.
\eeq 

Now let $x$ be a point  in $\S$ furthest from $\d\S$ in $\Om$, as in the 
definition of $R(\S)$.  Let $\g$ be a shortest curve in $(\S, \tilde h)$ from $x$
to $\d\S$.   Then $\g$ is a geodesic in $(\S, \tilde h)$, and by Synge's formula \cite{ON}
for the second variation of arc length, we have along $\g$,
\beq\label{ineq}
\int_0^{\tilde \ell} \left(\frac{df}{d\tilde s}\right)^2 - \tilde \kappa f^2\, d \tilde s \ge 0 \,,
\eeq
for all smooth functions $f$ defined on $[0,\tilde \ell]$ that vanish at the end points, where
$\tilde \ell$ is the $\tilde h$-length of $\g$ and $\tilde s$ is $\tilde h$-arc length along $\g$.
By making the change of variable $s  = s(\tilde s)$, where $s$ is $h$-arc length along $\g$,
and using Equation \eqref{gauss}, we arrive at,
\beq\label{ineq2}
\int_0^{\ell} \psi^{-1}(f')^2 - (Q + \psi^{-2} |\D\psi|^2)\psi^{-1}  f^2  \, d s \ge 0 \,,
\eeq
for all smooth functions $f$ defined on $[0,\ell]$ that vanish at the endpoints, where
$\ell$ is the $h$-length of $\g$, and $' = \frac{d}{ds}$.

Setting $k= \psi^{-1/2}f$ in \eqref{ineq2},  we obtain after a small
manipulation,
\beq\label{ineq3}
\int_0^{\ell}  (k')^2  -  Q\,k^2+ \psi^{-1}\psi'kk' -\frac34\psi^{-2}(\psi')^2k^2 \, ds \ge 0  \,,
\eeq
where $\psi'$ is shorthand for $(\psi \circ \g)'$, etc.
Completing the square on the last two terms of the integrand,
\beq
\frac34\psi^{-2}(\psi')^2k^2 - \psi^{-1}\psi'kk' = \left(\frac{\sqrt{3}}2 \psi^{-1}\psi'k -\frac1{\sqrt{3}} k'\right)^2 - \frac13(k')^2 ,  \nonumber
\eeq
we see that \eqref{ineq3} implies,
\beq\label{ineq4}
\int_0^{\ell}  \frac43(k')^2  -  Q\,k^2 \, ds \ge 0 \,.
\eeq
Since,  from \eqref{Q}, we have that  $Q \ge \mu - |J| \ge c$,
\eqref{ineq4}  implies,
\beq\label{ineq5} 
\frac43\int_0^{\ell}  (k')^2 \,ds \ge  c  \int_0^{\ell} k^2 \, ds \,.
\eeq
Setting $k = \sin \frac{\pi s}{\ell}$  in \eqref{ineq5} then gives,
\beq
\ell \le \frac{2\pi}{\sqrt {3}}\cdot \frac1{\sqrt{c}} \,.
\eeq 
Since $R(\S) \le \ell$, the result follows. \qed

\section{On the area of black holes in asymptotically anti-de Sitter spacetimes}

A basic step in the classical black hole uniqueness theorems is Hawking's theorem
on the topology of black holes \cite{HE} 
which asserts that cross sections of the event horizon
in $3+1$-dimensional asymptotically flat stationary black hole spacetimes obeying the
dominant energy condition are topologically 2-spheres.  As shown by Hawking \cite{Hawking},
this conclusion also holds for outermost MOTSs in spacetimes that
are not necessarily stationary.    In \cite{GS, G}  a natural
generalization of these results to higher dimensional spacetimes  was 
obtained by showing that cross sections of the event horizon (in the
stationary case) and outermost MOTSs (in the general case)
are of positive Yamabe type, i.e., admit metrics of positive
scalar curvature.  This implies many well-known restrictions on
the topology, and is consistent with recent examples of five
dimensional stationary black hole spacetimes with horizon topology
$S^2 \times S^1$ \cite{Emp}.  

These results on black hole topology depend crucially on the dominant
energy condition.  Indeed, there is a well-known class of  $3+1$-dimensional static locally anti-de Sitter black hole spacetimes which are solutions to the vacuum Einstein equations 
with negative cosmological constant $\Lambda$
having horizon topology of arbitrary genus $g$ \cite{Brill, Mann}.  Higher dimensional versions
of these topological black holes have been considered in \cite{Bir, Mann}.    However,
as Gibbons  pointed out in \cite{Gi}, although Hawking's theorem does not hold in the asymptotically locally anti-de Sitter setting,  his basic argument still
leads to an interesting conclusion.  Gibbons showed that for $3$-dimensional time-symmetric initial data sets that
give rise to spacetimes satisfying the Einstein equations with $\Lambda <0$,  
outermost MOTSs  $\S$ (which are  stable minimal surfaces in this case)  must satisfy the area bound,
\beq\label{areabound}
{\rm Area}(\S)\ge \frac{4\pi(g-1)}{|\Lambda|} \, ,
\eeq
where $g$ is the genus of $\S$.   Woolgar \cite{Wo}  obtained a similar bound in  the general, nontime-symmetric, case.
Hence, at least for stationary black holes, black hole entropy has a lower bound depending on a global topological invariant.

In \cite{CG} Cai and Galloway considered  an extension of Gibbon's result to higher dimensional spacetimes.  There  it was shown, for time-symmetric initial data, that a bound similar to that obtained by Gibbons still holds, but where the genus is replaced by the
so-called $\s$-constant (or Yamabe invariant).  The $\s$-constant is a diffeomorphism invariant of smooth compact manifolds that in dimension two reduces to a multiple of the 
Euler characteristic; see \cite{CG} and references therein for further details.  The
aim of this section is to observe that this result extends to the general, nontime-symmetric case.

We begin by recalling the definition of the $\s$-constant.  
Let $\S^{n-1}$, $n\ge 3$, be a smooth compact (without boundary) $(n-1)$-dimensional
manifold.  If $g$ is a Riemannian metric on $\S^{n-1}$, let $[g]$ denote the conformal
class of $g$.  The Yamabe constant with respect to $[g]$, which we denote by $\calY[g]$, is the number, 
\beq\label{yam}
\calY[g]  = \inf_{\tilde g\in [g]} 
\frac{\int_{\S}S_{\tilde g}d\mu_{\tilde g}}
{(\int_{\S}d\mu_{\tilde g})^{\frac{n-3}{n-1}}}\, ,
\eeq  
where $S_{\tilde g}$ and $d\mu_{\tilde g}$ are respectively the scalar curvature and volume measure of $\S^{n-1}$ 
in the metric $\tilde g$.  The  expression involving integrals is just the volume-normalized total
scalar curvature of $(\S,\tilde g)$.
The solution to the Yamabe problem, due to Yamabe, Trudinger, Aubin and Schoen, 
guarantees that the infimum 
in (\ref{yam}) is  achieved by a metric of constant scalar curvature.  

The $\s$-constant  of $\S$ is 
 defined by taking the supremum of the Yamabe constants over all conformal
classes,
\beq
\s(\S) = \sup_{[g]} \calY[g] \, .
\eeq    
As observed by Aubin, the supremum is finite, and in fact bounded above in terms of the volume
of the standard unit $(n-1)$-sphere $S^{n-1} \subset \Bbb R^n$.  The $\s$-constant divides
the family of compact manifolds into three classes according to: (1) $\s(\S) > 0$, (2) $\s(\S) = 0$,
and (3) $\s(\S) < 0$.  

In the case $\dim \S =2$, the Gauss-Bonnet theorem implies $\s(\S) = 4\pi\chi(\S)=8\pi(1-g)$.  
Note that the inequality \eqref{areabound} only gives information when $\chi(\S) < 0$.
Correspondingly, in higher dimensions, we shall only be interested in the case 
when $\s(\S) < 0$.  It follows from the resolution of the Yamabe problem that
$\s(\S) \le 0$ if and only if $\S$ does not carry a metric of positive scalar curvature.
In this case, and with $\dim \S = 3$, Anderson \cite{An} has shown, as an application of Perlman's work on
the geometrization conjecture, that $\s(\S)$  is 
determined by the volume of the ``hyperbolic part'' of $\S$, which when present
implies $\s(\S) < 0$.  In particular, all closed
hyperbolic $3$-manifolds have negative $\s$-constant.

We now turn to the spacetime setting.  In what follows, all MOTSs are compact
without boundary.  The following theorem extends Theorem 5 in \cite{CG}
to the nontime-symmetric case.  

\begin{thm}\label{volbound}  
Let $\S^{n-1}$ be a stable MOTS in the initial data set $(V^n,g,K)$, $n \ge 4$,
such that $\s(\S) < 0$.  Suppose there exists $c > 0$, such that  $\mu +\<J,\nu\> \ge -c$.
Then the $(n-1)$-volume of $\S$ satisfies,
\beq
{\rm vol}(\S^{n-1}) \ge \left(\frac{|\s(\S)|}{2c}\right)^{\frac{n-1}2} \, .
\eeq   
\end{thm}

We make some comments about the assumptions.  Suppose $V$ is a spacelike hypersurface in a spacetime $(M,g_M)$, satisfying the Einstein equation with cosmological
term,
\beq
G + \Lambda g_M = \calT
\eeq
where, as in Section 2, $G = \ric_M -\frac12R_Mg_M$ is  the Einstein tensor, and $\calT$
is the energy-momentum tensor. 
Thus,
setting $\ell = u+\nu$, we have along $\S$ in $V$,
\begin{align}
\mu +\<J,\nu\> &= G(u,\ell) = \calT(u,\ell) + \Lambda \nonumber \\
& \ge - |\Lambda| \,,
\end{align}
provided $\Lambda < 0$ and $\calT(u,\ell) \ge 0$.  Hence, when $\Lambda < 0$ and the fields giving rise to $\calT$ obey the dominant energy condition, the energy condition in Theorem \ref{volbound} is satisfied with $c = |\Lambda|$.   

We briefly comment on the stability assumption. As defined in \cite{G}, a MOTS $\S$ is {\it weakly outermost} in $V$ provided there are no strictly outer trapped surfaces  outside of, and  homologous to $\S$ in $V$.  Weakly outermost MOTSs are necessarily stable, as noted in Section 2, 
and arise naturally in a number of physical
situations.  For example,  smooth compact cross sections of the
event horizon in 
stationary black hole spacetimes obeying the null energy condition, are 
necessarily weakly outermost MOTSs.
Moreover, results of Andersson and Metzger \cite{AM2} provide natural criteria for the existence of weakly outermost MOTSs in general black hole spacetimes containing
trapped regions. 

\proof[Proof of Theorem \ref{volbound}] The proof is a simple  modification of the
proof of Theorem 5 in \cite{CG}.
By the stability assumption and Proposition
\ref{eigen}, we have $\l_1(L_0) \ge 0$, where $L_0$ is the operator  given in  \eqref{symop}. 
The Rayleigh formula,
$$
\l_1(L_0) = \inf_{\phi \ne 0} \frac{ \int_{\S} \phi L_0(\phi)d\mu}{\int_{\S} \phi^2 d\mu}
$$ 
together with an integration by parts yields the {\it stability inequality},
\beq\label{stab}
\int_{\S} (|\D \f|^2 + \left( \frac12 S - (\mu + \<J,\nu\>) - \frac12 |\chi|^2\right)\phi^2 \, d\mu \ge 0 \, ,
\eeq
for all $\phi \in C^{\infty}(\S)$.

The Yamabe constant $\calY[h]$, where $h$ is the induced metric on $\S$, can
be expressed as \cite{Be},
\beq\label{yam2}
\calY[h] = \inf_{\f\in C^{\8}(\S), \f>0} \frac{\int_{\S} (\frac{4(n-2)}{n-3}|\D\f|^2 +S\f^2)\, d\mu}
{(\int_{\S}\f^{\frac{2(n-1)}{n-3}}\, d\mu)^{\frac{n-3}{n-1}}} \, .
\eeq

Noting that $\frac{4(n-2)}{n-3}> 2$, the stability inequality implies,
\begin{align}\label{ineqc}
\int_{\S} \frac{4(n-2)}{n-3}|\D \f|^2 + S\f^2) \, d\mu & \ge  \int_{\S}
2( \mu + \<J,\nu\> )\phi^2\,d\mu
\nonumber\\
& \ge  - 2c \int_{\S} \f^2\,d\mu \, .
\end{align}

By H\"older's inequality we have,
\beq
\int_{\S} \f^2\,d\mu \le \left(\int_{\S} \f^{\frac{2(n-1)}{n-3}}\, d\mu\right)^{\frac{n-3}{n-1}}
\left(\int_{\S} 1 \, d\mu\right)^{\frac2{n-1}}  \, ,
\eeq
which, when combined with  (\ref{ineqc}), gives,
\beq
\frac{\int_{\S} (\frac{4(n-2)}{n-3}|\D\f|^2 + \hat S\f^2)\, d\mu}
{(\int_{\S}\f^{\frac{2(n-1)}{n-3}}\, d\mu)^{\frac{n-3}{n-1}}} \ge - 2c\,(\mbox{vol($\S$)})^{\frac2{n-1}} \, .
\eeq
Making use of this  inequality in (\ref{yam2}) gives,
$\calY[h] \ge - 2c\, (\mbox{vol($\S$)})^{\frac2{n-1}}$, or, equivalently,
\beq
{\rm vol}(\S^{n-1}) \ge \left(\frac{|\calY[h]|}{2c}\right)^{\frac{n-1}2} \, .
\eeq  
Since $|\s(\S)| \le |\calY[h]|$, the result follows.\qed

\section*{Acknowledgements}

\vspace{-.1in}
This work was supported in part by NSF grant
DMS-0708048 (GJG) and SFI grant 07/RFP/PHYF148 (NOM).


\providecommand{\bysame}{\leavevmode\hbox to3em{\hrulefill}\thinspace}
\providecommand{\MR}{\relax\ifhmode\unskip\space\fi MR }
\providecommand{\MRhref}[2]{%
  \href{http://www.ams.org/mathscinet-getitem?mr=#1}{#2}
}
\providecommand{\href}[2]{#2}

\end{document}